\documentclass{PoS}

\usepackage[T1]{fontenc} 
\usepackage{graphicx}
\usepackage{graphics}
\usepackage{subfig}
\usepackage[mathscr]{eucal}
\usepackage{amssymb}
\usepackage{amsmath}
\usepackage{xspace}
\usepackage{listings}
\usepackage{soul}
\usepackage{ulem}
\usepackage{color}
\usepackage[dvipsnames]{xcolor}
\usepackage{ulem}
\usepackage{bm}         
\usepackage{array}
\usepackage{multirow}   
\usepackage{booktabs}   
\usepackage{mleftright}
\usepackage[utf8]{inputenc}
\usepackage{caption}

\renewcommand{\(}{\left(}
\renewcommand{\)}{\right)}

\newcommand{\U}[1]{\mathrm{U}(1)_{\mathrm{#1}}}			

\title{Multi-peaked signatures of primordial gravitational waves from multi-step electroweak phase transition}

\ShortTitle{Multi-peak gravitational wave signatures}

\author{Ant{\'o}nio~P.~Morais\\
        Departamento de F\'isica, Universidade de Aveiro and CIDMA, Campus de Santiago, 3810-183 Aveiro, Portugal\\
        E-mail: \email{a.morais.physics@gmail.com}}

\author{\speaker{Roman~Pasechnik}\\
        Department of Astronomy and Theoretical Physics, Lund University, 221 00 Lund, Sweden\\
        E-mail: \email{Roman.Pasechnik@thep.lu.se}}

\author{Thibault~Vieu\\
        Magist\`ere de Physique Fondamentale, Universit\'e Paris-Saclay, B\^at. 470, F-91405 Orsay, France\\
        International Centre for Fundamental Physics, Ecole Normale Sup\'erieure,  24 rue Lhomond, 75005 Paris, France\\
        E-mail: \email{vieu@apc.in2p3.fr}}

\abstract{The first-order electroweak phase transition in the early universe could occur in multiple steps leading to specific multi-peaked signatures 
in the primordial gravitational wave (GW) spectrum. We argue that these signatures are generic phenomena in multi-scalar extensions of 
the Standard Model. In a simple example of such an extension, we have studied the emergence of reoccurring and nested vacuum bubble 
configurations and their role in the formation of multiple peaks in the GW spectrum. The conditions for potential detectability of these 
features by the forthcoming generation of interferometers have been studied.
}

\FullConference{
European Physical Society Conference on High Energy Physics - EPS-HEP2019 -\\
			10-17 July, 2019\\
			Ghent, Belgium}

\begin{document}

\noindent \textbf{\textit{Introduction.}}~Despite of the great success of measurements at the Large Hadron Collider (LHC), the persistent absence of new physics evidence 
is driving an increasing discomfort among the particle physics community. 
The current void of new phenomena either indicates that new physics can only be manifest at a larger energy scale than previously thought, or results from a lack of 
sensitivity of the current experiments measuring rare events. In fact, the weaker the interaction strength between the SM and new physics, the greater the challenge 
to probe it.

On the other hand, the recent discovery of a binary neutron star merger, firstly observed by the gravitational waves (GW) interferometers of the LIGO-Virgo collaboration 
\cite{Abbott:2016blz,Abbott:2016nmj}, a new era of multi-messenger astronomy has begun. Furthermore, the reach of GW observatories 
is by no means exhausted and larger sensitivities are designed for future space-based interferometers such as those of the LISA \cite{Bartolo:2016ami}, 
DECIGO \cite{Kawamura:2011zz} and BBO \cite{Corbin:2005ny} collaborations. This opens up the door for a plethora of new studies including connections 
with both cosmology and particle physics (see e.g.~Refs.~\cite{Morais:2019fnm,Addazi:2019dqt} and references therein). In particular, the potential observation 
of a stochastic GW background produced by violent processes in the early universe, e.g.~by expanding vacuum bubbles associated with strong cosmological 
phase transitions \cite{Hindmarsh:2013xza,Hindmarsh:2015qta}, may well become a gravitational probe for beyond-the-SM (BSM) physics and a complement for collider measurements.
When extending the SM scalar sector, the underlying vacuum structure exhibits a growing complexity such that a possibility for transition patterns with several successive 
first-order steps arises~\cite{Morais:2019fnm,Addazi:2019dqt,Patel:2012pi,Inoue:2015pza,Vaskonen:2016yiu}. In this contribution, we discuss the key 
implications of such successive transitions for GW signals.
\par
\noindent \textbf{\textit{Multi-step electroweak phase transition.}}~In cosmology, thermal evolution of the EW-breaking vacuum as the universe cools down is determined 
by the temperature-dependent part of the one-loop effective potential \cite{Quiros:1999jp}. Given the field content and its quantum numbers of an underlying multi-scalar fundamental 
theory, the shape of the potential can be determined at any temperature $T$. In a configuration of two minima of the effective potential coexisting at the critical temperature $T_c$, 
by using CosmoTransitions~\cite{Wainwright:2011kj} one computes numerically the Euclidean action $\hat{S}_3$ describing transitions 
between the corresponding phases. The temperature $T_n$, at which the nucleation of vacuum bubbles effectively occurs, is estimated by setting the probability to nucleate 
one bubble per horizon volume to unity, which translates into $\hat{S}_3 / T_n \sim 140$ \cite{Dine:1992wr,Quiros:1999jp}. The sphaleron suppression criterion 
$v_c/T_c \gtrsim 1$, with $v_c=v(T_c)$ the Higgs vacuum expectation value (VEV), defines a strong first-order phase transition. This typically implies 
that the transition produces strong GW signals detectable by the next generation of interferometers and may be used for constraining BSM scenarios 
\cite{Kakizaki:2015wua,Hashino:2016rvx,Hashino:2016xoj}. Here we study successive strong first-order EWPTs which we refer to as multi-step transitions. 
As a result one has more than a single transition pattern for a particular point in the parameter space, which results in sequential nucleation of bubbles of different vacua. 

A generic BSM scenario typically contains a large number of scalar degrees of freedom which can be advantageous e.g.~for EW baryogenesis. Even reducing the scalar 
sector to a few fields, new unexplored possibilities of transition patterns arise, in particular, transitions in several successive first-order steps. Therefore, a non-trivial EWPT is expected 
and multi-step transitions may have occurred in the early universe \cite{Morais:2019fnm}. The basic characteristics of such multi-peak spectra may also be 
affected by dynamics of other sectors, in particular, by the neutrino sector suggesting the use of primordial GW data for probing the neutrino mass generation
mechanisms~\cite{Addazi:2019dqt}.

In order to illustrate the generic features of multi-step first-order EW transitions, consider for instance a minimal extension of the SM scalar sector inspired by 
the high-scale Grand-unified trinification theory \cite{Camargo-Molina:2016yqm,Camargo-Molina:2017kxd,Vieu:2018nfq}. Besides the SM Higgs field 
$\mathcal{H}_1$, it contains an additional EW doublet $\mathcal{H}_2$ and singlet $\varphi$ fields which are charged under a $\U{}$ family symmetry. 
The resulting potential possesses an approximate descrete $\mathbb{Z}_2$ symmetry acting as $\mathcal{H}_j \to - \mathcal{H}_j$ ($j=1,2$) and 
$\varphi \to - \varphi$ which significantly simplifies the vacuum structure of the model. An expansion of the scalar fields in terms of real components
\begin{align}
\begin{aligned}
\mathcal{H}_j &= \frac{1}{\sqrt{2}} \begin{pmatrix} \chi_j + i \chi_j' \\ \phi_j + h_j + i \eta_j \end{pmatrix}\,,
\end{aligned} \,\,
\begin{aligned}
\varphi = \dfrac{1}{\sqrt{2}} \( \phi_s + S_R + i S_I\)\,,
\end{aligned}
\end{align}
defines the quantum fluctuations $h_1$, $h_2$ and $S_R$ about the classical configurations $\phi_\alpha=\{\phi_1,\phi_2,\phi_s\}$, respectively. With this expansion, 
the classical potential reads
\begin{align}
\label{eq:2HDMS}
\begin{aligned}
V_{\rm cl}(\phi_\alpha) =& 
  \tfrac{1}{2} m_\alpha^2 \lvert\phi_\alpha\rvert^2
+ \tfrac{1}{8} \lambda_\alpha \lvert\phi_\alpha\rvert^4 
+ \tfrac{1}{4} \lambda_{\alpha\beta} \lvert \phi_\alpha\rvert^2 \lvert \phi_\beta\rvert^2\,.
\end{aligned}
\end{align}
A comprehensive analysis of the tree-level vacuum structure was performed recently in Ref.~\cite{Vieu:2018nfq}. It was shown that the basic characteristics of 
EWPTs in this model, in particular, sequential first-order transitions, are generic for multi-Higgs extensions of the SM. Thus, this model, due to 
the simplicity of its potential \eqref{eq:2HDMS}, could serve as a good benchmark scenario for further in-depth explorations of cosmological implications of 
multi-scalar BSM theories.
\begin{figure}[!h]
\begin{minipage}{0.30\textwidth}
 \centerline{\includegraphics[width=1.0\textwidth]{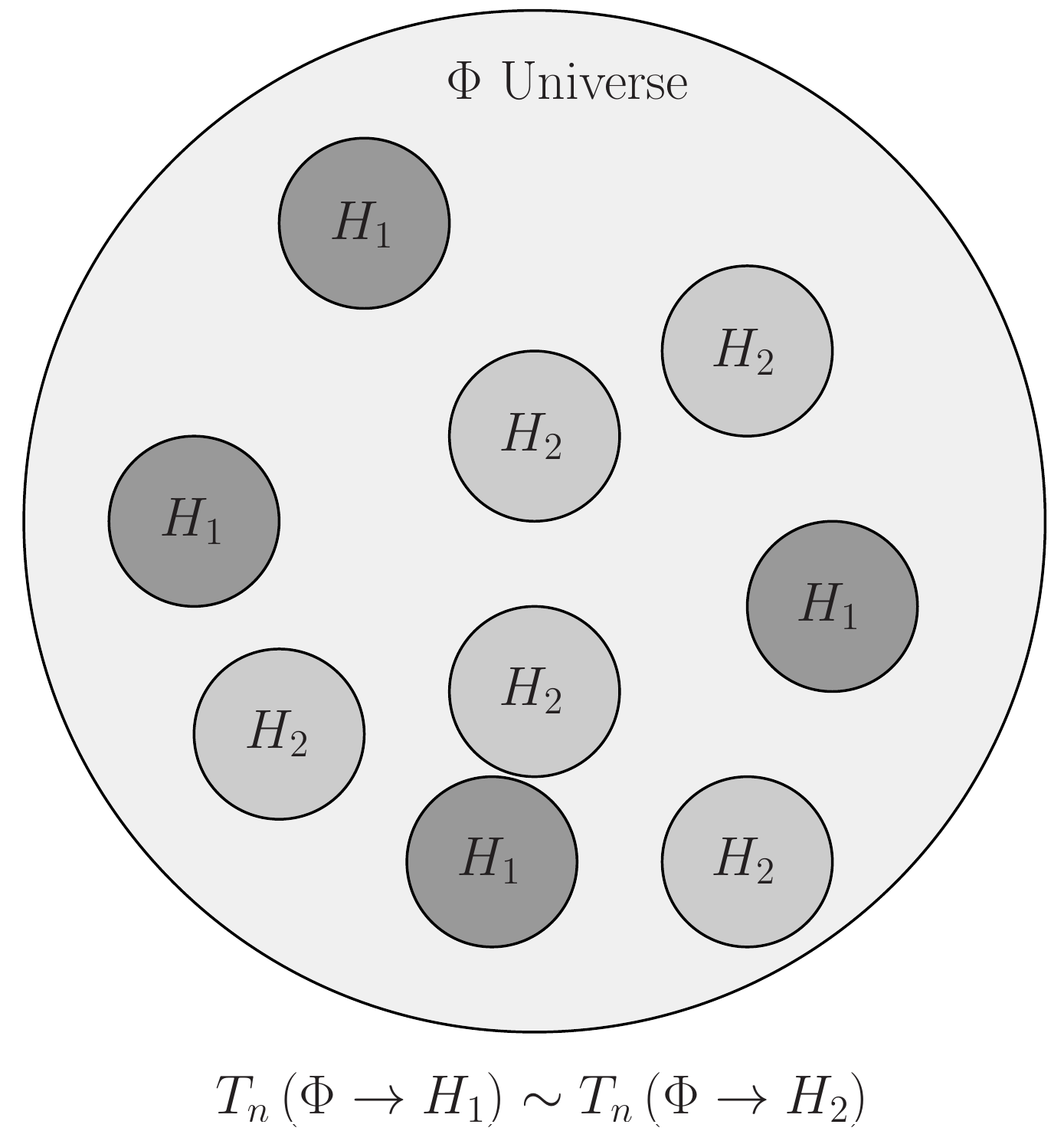}}
\end{minipage}
\hspace{0.6cm}
\begin{minipage}{0.30\textwidth}
 \centerline{\includegraphics[width=1.0\textwidth]{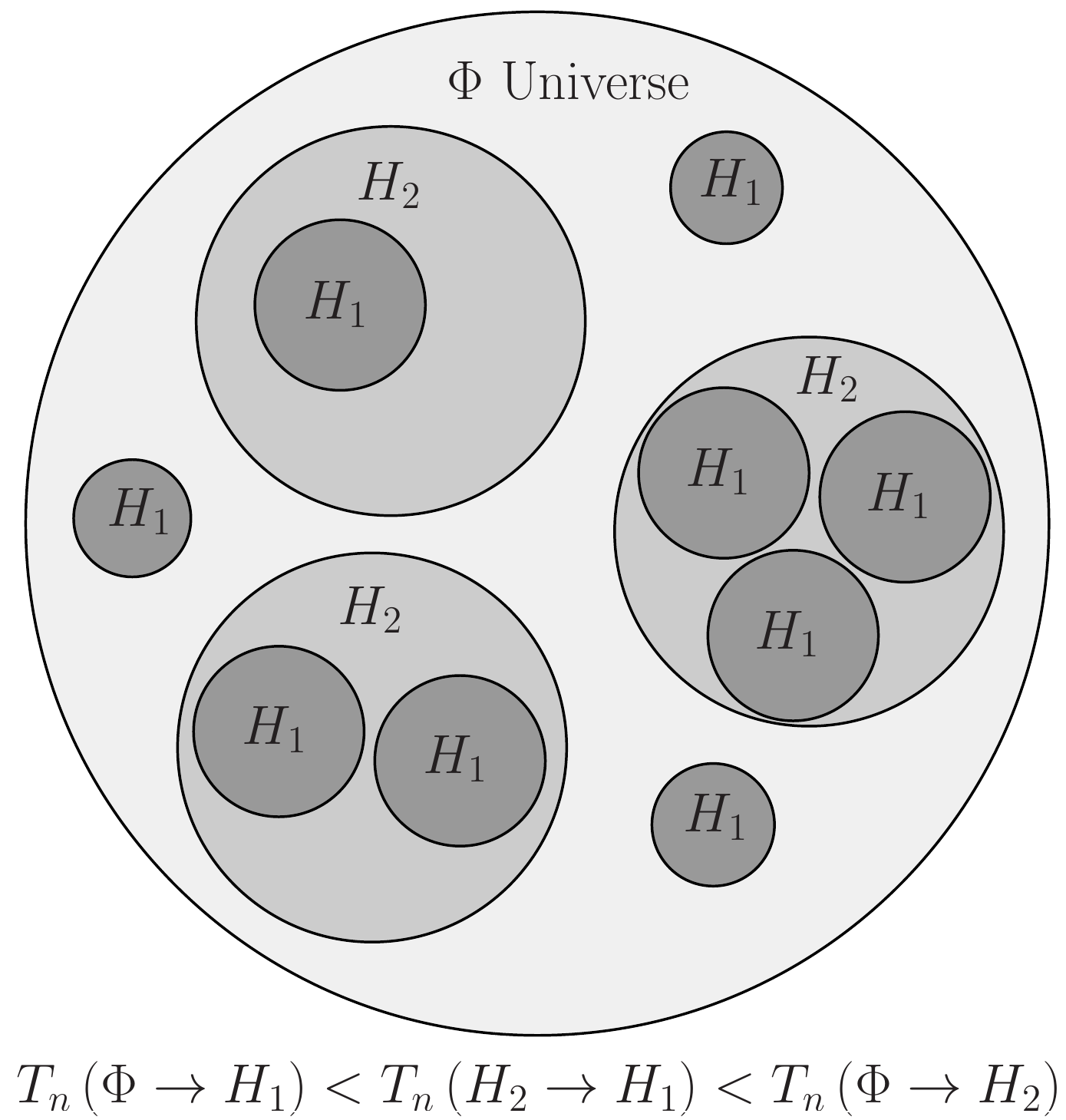}}
\end{minipage}
\hspace{0.6cm}
\begin{minipage}{0.30\textwidth}
 \centerline{\includegraphics[width=1.0\textwidth]{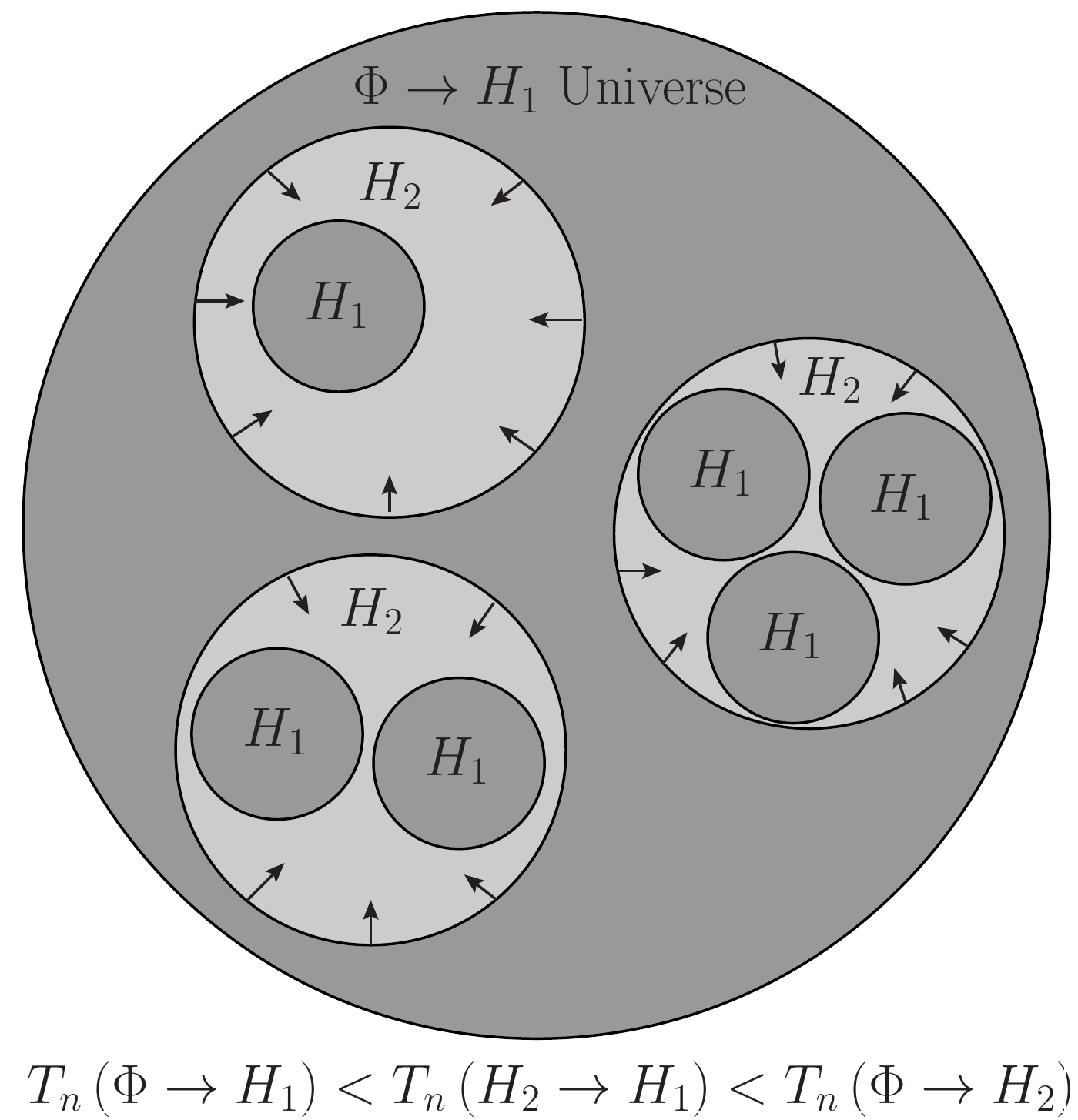}}
\end{minipage}
   \caption{
An illustration of a universe in the $\Phi$-phase filled with coexisting bubbles of new $H_1$ and $H_2$ phases emerging simultaneously (left panel), and also with 
nested bubbles when $H_1$-bubbles are born inside of $H_2$ ones (middle panel). In the right panel, the nucleation of the $H_1$-bubbles
in the $\Phi$ phase causes the previously produced $H_2$-bubbles to contract simultaneously with nucleation of smaller $H_1$-bubbles inside 
them (reoccuring bubbles).
}
\label{fig:bubbles}
\end{figure}

\par
\noindent \textbf{\textit{Coexisting, nested and reoccurring bubbles.}}~For simplicity, let us now consider a representative configuration of the parameter space \cite{Vieu:2018nfq} 
where the only existing phases given in terms of the VEVs of the scalar fields $v_\alpha\equiv\langle \phi_\alpha\rangle_{\rm vac}=\{v_1,v_2,v_s\}$ are $(0,0,0)$, $(v_1,0,0)$, $(0,v_2,0)$ 
and $(0,0,v_s)$, which we recast as $[0]$, $H_1$, $H_2$ and $\Phi$, respectively. The possible first-order transitions were found to be $H_1 \leftrightarrow H_2$, $H_1 \leftrightarrow \Phi$, 
$H_2 \leftrightarrow \Phi$, which take place readily in the leading $(m/T)^2$ order of the thermal expansion and are expected to be strong. Here, without loss 
of generality we identify the stable phase at $T=0$ with $H_1$ which is simple but represents the basic features of a generic 
EW-breaking vacuum $\{v_1,v_2,0\}$.

In a particular sequence of transitions to the true vacuum $H_1$ with the following two patterns
\begin{equation}
\label{Eq:pat1}
\Phi \to H_1 \,, \qquad \Phi \to H_2 \to H_1\,,
\end{equation}
one could expect several nucleation processes occurring in the same range of temperatures, e.g.~$\Phi \to H_1$ and $\Phi \to H_2$. In this case, different sequences 
could be realized during the same cosmological evolution time leading to a universe where {\it coexisting bubbles} of different broken phases expand simultaneously 
(left panel in Fig.~\ref{fig:bubbles}). In addition to the coexisting bubbles, more exotic cosmological objects may emerge from multi-step phase transitions. In particular, 
consider the second and third steps in the pattern $[0] \to \Phi \to H_2 \to H_1$, occurring at typical nucleation temperatures $T_n(\Phi \to H_2) \gtrsim T_n(H_2 \to H_1)$. 
Between $T_n(\Phi \to H_2)$ and $T_n(H_2 \to H_1)$, the $H_2$-bubbles nucleate and expand in a universe filled with the $\Phi$-phase. Then at $T_n(H_2 \to H_1)$, while 
they are still expanding, the $H_1$-bubbles emerge and nucleate inside the $H_2$-bubbles. As such, the $\Phi$-phase becomes populated with the $H_2$-bubbles 
containing the $H_1$-bubbles inside. We denote such objects as {\it nested bubbles} shown in Fig.~\ref{fig:bubbles} (middle panel).
\begin{figure}
\centering
\includegraphics[width=0.5\linewidth]{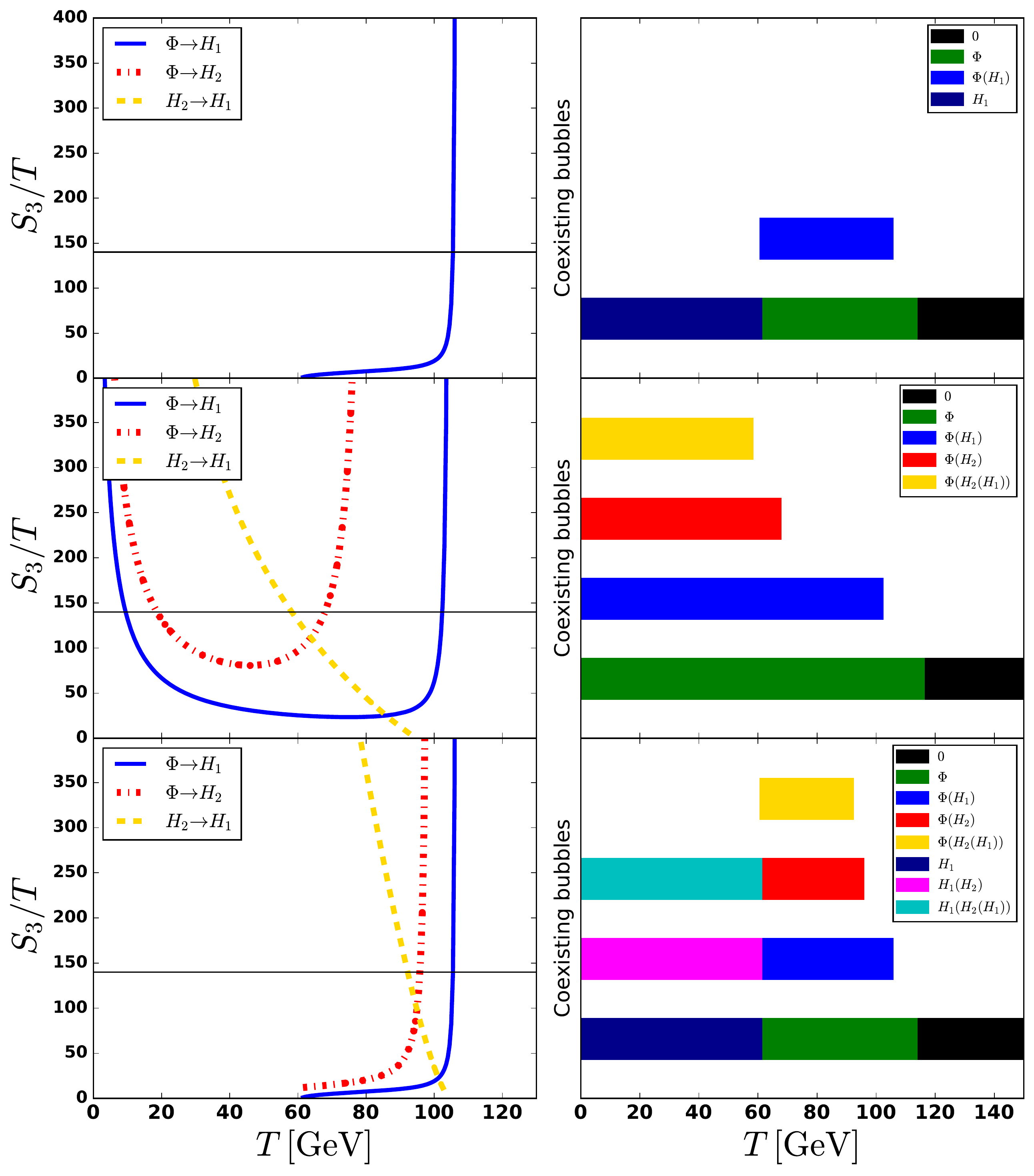}
\caption{\label{fig:exotic_processes} 
Left column: evolution of $\hat{S_3}/T$ for all possible transitions. 
Right column: diagrams representing all types of bubbles (co)existing at a given temperature.}
\end{figure}

Since the scalar potential keeps evolving as the universe cools down below $T_n(H_2 \to H_1)$, the initial phase $\Phi$ becomes unstable also along the $H_1$ direction, 
so both transitions towards $H_1$ and $H_2$ can occur (coexisting bubbles scenario). In particular, if the potential barrier between the phases $\Phi$ and $H_1$ disappears, 
the new $H_1$-bubbles would nucleate in the parts of the universe that still remain in the $\Phi$-phase i.e.~the direct $\Phi \to H_1$ transition quickly eliminates 
the $\Phi$-phase outside of the $H_2$-bubbles formed at an earlier time. Such a mixed situation with the coexistence of the ordinary $H_1$ and nested $H_2 \to H_1$ 
bubbles is depicted in Fig.~\ref{fig:bubbles} (middle panel). In the end of this process, one ends up with the $H_1$-bubbles inside the $H_2$-bubbles which exist in a universe 
filled with the $H_1$-phase. We denote these exotic cosmological objects as {\it reoccurring bubbles}. Since the $H_2$-bubbles cannot expand in a universe filled with 
the stable $H_1$-phase, they are pushed inwards and collapse while the $H_1$-bubbles nucleate inside them as illustrated in Fig.~\ref{fig:bubbles} (right panel).

In Fig.~\ref{fig:exotic_processes} we show three realistic cosmological scenarios where the objects discussed above are expected to occur. In the left column, 
we plot the evolution of the action $\hat{S}_3/T$ as a function of temperature for all possible transitions. Whenever a curve corresponding to a transition $i \to j$ 
crosses the horizontal line $\hat{S}_3/T = 140$, a bubble of phase $j$ is nucleated inside the phase $i$, which is denoted as $i(j)$ in what follows. In the right column, 
we show a diagrammatic representation displaying all types of bubbles (co)existing at a given temperature and corresponding to the plots in the left 
column. For example, the top panel describes a first-order phase transition $\Phi \to H_1$ when $H_1$-bubbles nucleate in a universe filled with 
the $\Phi$-phase (i.e.~$\Phi(H_1)$). In particular, a universe in the symmetric phase $[0]$ first collapses to the $\Phi$-phase through a second-order 
phase transition without generating any bubbles. Then, the $H_1$-bubbles are nucleated at $T\sim 105$~GeV and expand until the $\Phi$-phase becomes 
unstable at around $T\sim 60$ GeV leaving a universe entirely filled by the true vacuum $H_1$.
\begin{figure}[!h]
\begin{minipage}{0.495\textwidth}
 \centerline{\includegraphics[width=1.0\textwidth]{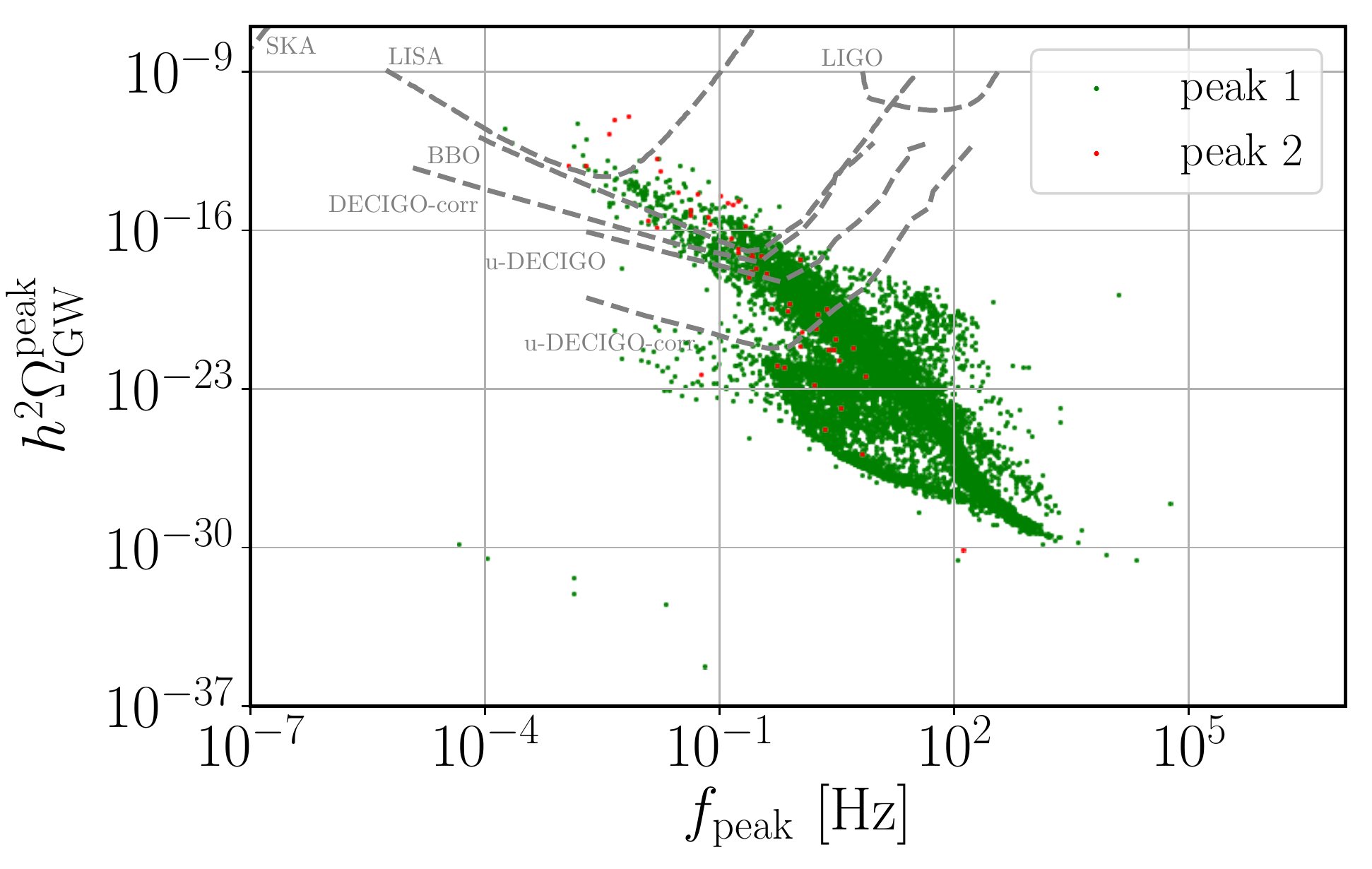}}
\end{minipage}
\begin{minipage}{0.45\textwidth}
 \centerline{\includegraphics[width=1.0\textwidth]{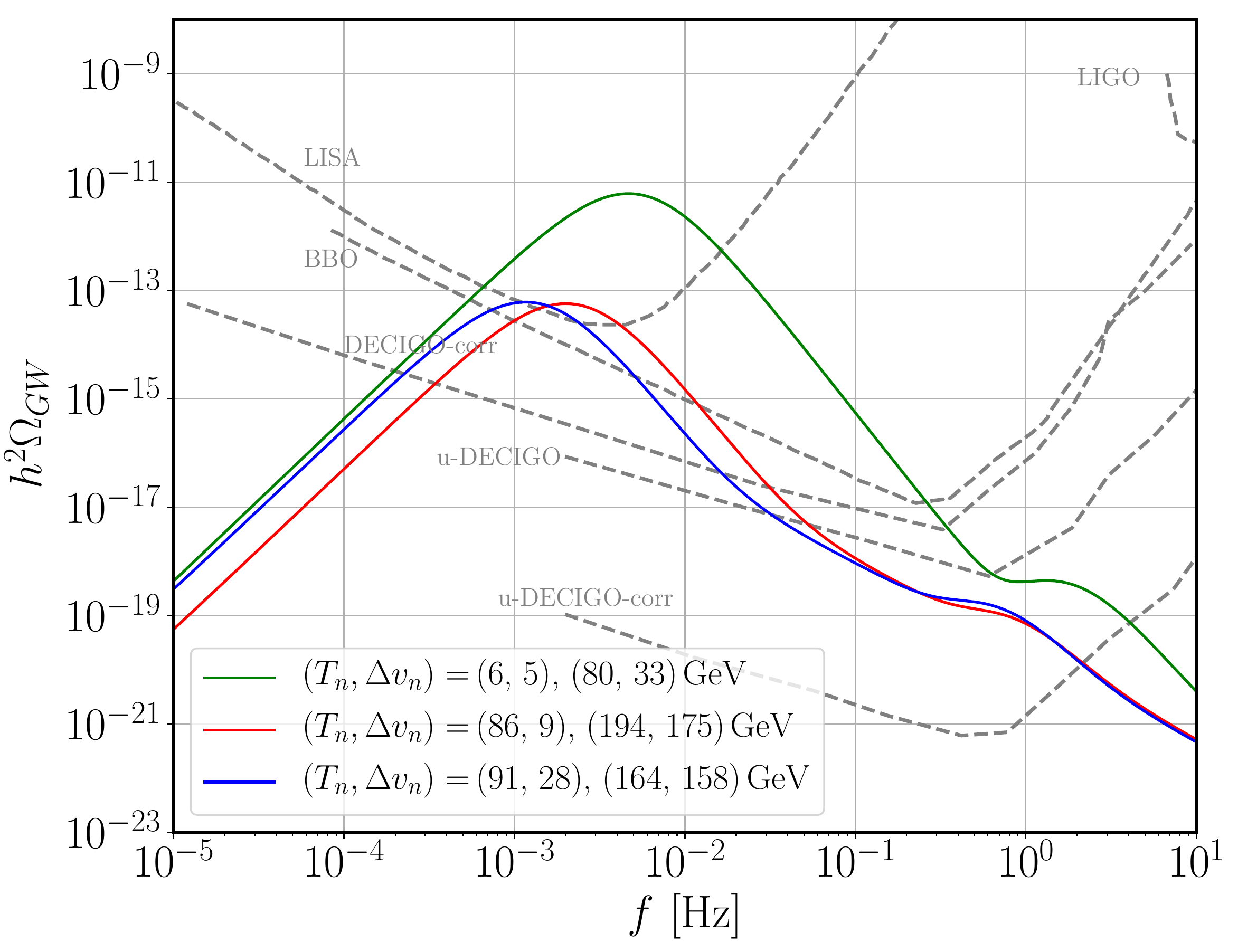}}
\end{minipage}
   \caption{
{\bf Left panel}: The results of the inclusive parameter scan in the considering 2HDSM where each point corresponds to a FOPT with 
the corresponding GW peak amplitude and frequency. Double-peak signatures are shown by red dots.
{\bf Right panel}: Benchmark double-peak GW spectra with the largest peaks entering the sensitivity domain 
of the LISA experiment with main characteristics specified in Table~\ref{tab:properties_GW}.
}
\label{fig:GW}
\end{figure}

While the graphs on the second line of Fig.~\ref{fig:exotic_processes} represent a scenario where three successive steps lead to the nucleation of a nested 
bubbles $\Phi(H_2(H_1))$, those on the last line describe reoccurring bubbles that emerges from a nested ones. For instance, after nucleation of nested 
bubbles $\Phi(H_2(H_1))$, the potential barrier between the phases $\Phi$ and $H_1$ disappears (around $T\sim 60$ GeV), such that the parts of 
the universe in the $\Phi$-phase collapse to the $H_1$-phase transforming the nested bubbles $\Phi(H_2(H_1))$ into the reoccurring ones $H_1(H_2(H_1))$. 
This type of cosmological objects is only possible if the nucleation temperatures of the corresponding steps are not too different (percolation typically occurs 
in the range of $\Delta T$ < 10 GeV), and further likely to occur when symmetries in the potential enforces them to be identical as in e.g.~Ref.~\cite{Ivanov:2017zjq}. 
\par
\noindent \textbf{\textit{Multi-peaked gravitational-wave spectrum.}}~We analyse the multi-peaked signatures in the power spectrum of GWs using the well known 
formalism of Ref.~\cite{Grojean:2006bp,Leitao:2015fmj,Caprini:2015zlo} 
which describes the energy density per logarithmic frequency of the GW radiation, $h^2 \Omega_{\rm GW}$. The net GW signal is typically considered to be produced by three different sources due to bubble wall collisions \cite{Jinno:2016vai}, sound waves (SW) generated by the phase transitions \cite{Hindmarsh:2013xza}, 
as well as magnetohydrodynamics (MHD) turbulences in the plasma \cite{Caprini:2009yp}. The bubble wall collisions typically do not contribute to the GWs production 
processes in the considering class of multi-scalar extensions of the SM~\cite{Hindmarsh:2017gnf,Ellis:2019oqb} (see also Ref.~\cite{Morais:2019fnm}) so we account
for the dominant SW and MHD contributions only.
\begin{table}
\centering
\resizebox{\columnwidth}{!}{%
\begin{tabular}{|c|c|ccccc|cccccc|cc|} \hline
line &   $T_n$ &   $\alpha$ & $\beta/H$ &  $v_b$  & $v^i_1$ & $v^f_1$ & $v^i_2$ & $v^f_2$ 
       & $v^i_s$ & $v^f_s$ & $f_{\rm peak}$ & $h^2\Omega_{\rm GW}^{\rm peak}$ \\ \hline
green &  80 &  9.6$\cdot 10^{-3}$ & 6.1$\cdot 10^{4}$ & 0.65 & 0        & 87      & 86     & 81  & 0     & 0   & 1.6                       & 4$\cdot 10^{-19}$ \\
   &  6   &  0.4                          & 3.4$\cdot 10^{3}$ & 0.88 & 240    & 246   & 23     & 0     & 0    & 0   & 5$\cdot 10^{-3}$ & 6.1$\cdot 10^{-12}$ \\ \hline
red &  194 &  7.1$\cdot 10^{-3}$ & 1.1$\cdot 10^{4}$ & 0.64 & 0        & 0       & 0        & 175  & 0   & 0   & 0.6                      & 6.8$\cdot 10^{-20}$ \\
   &  86   &  0.1                          & 96                           & 0.79 & 0        & 240   & 231    & 0      & 0   & 0   & 2$\cdot 10^{-3}$ & 5.7$\cdot 10^{-14}$ \\ \hline
blue &  164 &  6.6$\cdot 10^{-3}$ & 1.1$\cdot 10^{4}$  & 0.64  & 0        & 0       & 0        & 158  & 0    & 0  & 0.5                     & 1.1$\cdot 10^{-19}$ \\
   &  91   &  8.6$\cdot 10^{-2}$ & 51                           &  0.77  & 0        & 235   & 207    & 0      & 0    & 0  & 1.3$\cdot 10^{-3}$ & 6$\cdot 10^{-14}$ \\ \hline
\end{tabular}%
}
\caption{Characteristics of selected benchmark double phase transitions whose GW signals emerge in the sensitivity domains
of planned measurements and whose GW spectra are illustrated in Fig.~\ref{fig:GW} by green, red and blue lines, respectively. 
Here, the nucleation temperature, $T_n$, the scalar VEVs before $v^i_\alpha$ and after $v^f_\alpha$ the respective phase transition 
are given in units of GeV, while the peak-frequency, $f_{\rm peak}$, is given in Hz. 
}
\label{tab:properties_GW}
\end{table}

The results of the inclusive parameter scan searching for FOPTs in the considering 2HDSM scenario are illustrated in Fig.~\ref{fig:GW} (left panel).
Here, each point corresponds to a particular FOPT found in the scan, with calculated values of the induced GW peak amplitude and frequency.
Scenarios corresponding to double-peak GW spectra are highlighted by red color. Here, dashed grey lines indicate sensitivities 
of the LISA \cite{Bartolo:2016ami} and LIGO \cite{Abbott:2016blz,Abbott:2016nmj} interferometers, 
as well as proposed DECIGO \cite{Kudoh:2005as,Kawamura:2011zz}, BBO \cite{Crowder:2005nr,Corbin:2005ny} 
and SKA \cite{Bull:2018lat} missions (see also Ref.~\cite{Buonanno:2004tp}). The ``ultimate-DECIGO'', ``ultimate-DECIGO-corr'' 
and ``DECIGO-corr'' sensitivity curves are taken from Ref.~\cite{Nakayama:2009ce}, while the sensitivities for other measurements
can be found in Ref.~\cite{Moore:2014lga}. We have selected three example scenarios whose largest peaks fall into the LISA sensitivity
domain and whose GW spectra are shown in Fig.~\ref{fig:GW} (right panel). The basic characteristics of the corresponding transitions
and GW signals for each green, red and blue line are summarised in Table~\ref{tab:properties_GW}. Such scenarios can be further 
considered as benchmarks for further explorations at GW interferometers. For a more detailed description of these and other benchmark
scenarios, see Ref.~\cite{Morais:2019fnm}.

We would like to point out that a complete knowledge of the bubble dynamics is needed in order to precisely describe the phase transitions, 
from nucleation to percolation. A few immediate questions result from our analysis: do we expect new sources of GWs when nested 
bubbles collide or when a nested bubble expands faster than its ``mother bubble'' reaching the wall of the latter? Given that in general such 
objects have no reason to be spherically symmetric, what is their impact on the profile of the GW spectrum? What is the impact of nested 
vacuum bubbles for baryogenesis? These are important questions for a further deeper analysis.
\acknowledgments
A.P.M.~is supported by Funda\c{c}\~ao 
para a Ci\^encia e a Tecnologia (FCT), within project UID/MAT/04106/2019 (CIDMA) and by national funds (OE), 
through FCT, I.P., in the scope of the framework contract foreseen in the numbers 4, 5 and 6 of the article 23, 
of the Decree-Law 57/2016, of August 29, changed by Law 57/2017, of July 19.~A.P.M.~is also supported by 
the \textit{Enabling Green E-science for the Square Kilometer Array Research Infrastructure} (ENGAGESKA), 
POCI-01-0145-FEDER-022217, and by the project \textit{From Higgs Phenomenology to the Unification 
of Fundamental Interactions}, PTDC/FIS-PAR/31000/2017.~R.P.~is supported in part by the Swedish 
Research Council grants, contract numbers 621-2013-4287 and 2016-05996, by the Ministry of Education, 
Youth and  Sports of the Czech Republic, project LT17018, as well as by the European Research Council (ERC) 
under the European Union's Horizon 2020 research and innovation programme (grant agreement No 668679). 
T.~V.~received a financial support Erasmus+ for international mobility from Universit\'e Paris-Sud.

\bibliographystyle{spphys}
\bibliography{bib}

\end{document}